\documentclass[traditabstract]{aa} 
\usepackage{graphicx}
\usepackage{txfonts}
\begin{document}
   \title{Corrigendum to``The upper atmosphere of the exoplanet HD209458b revealed by the sodium D lines:
Temperature-pressure profile, ionization layer and thermosphere''  
}

   \subtitle{[2011, A\&A, 527, A110]}

   \author{A.~Vidal-Madjar,
          \inst{1}
          C.M~Huitson,\inst{2}
          A.~Lecavelier des Etangs,\inst{1}
          D.K.~Sing,\inst{2}
          R.~Ferlet,\inst{1}
          J.-M.~D\'esert\inst{3}, 
          G.~H\'ebrard\inst{1}, 
          I.~Boisse\inst{1}, 
          D.~Ehrenreich\inst{4}, 
          \and 
          C.~Moutou\inst{5}
          }

   \institute{CNRS, UPMC, Institut d'astrophysique de Paris, UMR 7095, 98bis boulevard Arago 75014 Paris, France \\
              \email{alfred@iap.fr} 
         \and  Astrophysics Group, School of Physics, University of Exeter, Stocker Road, Exeter EX4 4QL, UK
         \and Harvard-Smithsonian Center for Astrophysics, 60 Garden St., Cambridge, MA 02138, USA
         \and Laboratoire dÕAstrophysique de Grenoble, Universit«e Joseph Fourier, CNRS (UMR 5571), BP 53, 38041 Grenoble cedex 9, France
         \and Laboratoire dÕAstrophysique de Marseille, Technopöole Marseille « Etoile, 38 rue Fr«ed«eric Joliot Curie, 13013 Marseille, France
             }

   \date{Received; accepted}

 \abstract
 {An error was detected in the code used for the analysis of the HD209458b sodium profile (Vidal-Madjar et al.,~2011). Here we present an updated $T$-$P$ profile and briefly discuss the consequences.}
 {}{}{}{}



\titlerunning{Corrigendum to HD209458b $T$-$P$ Profile}
\authorrunning{Vidal-Madjar et al., 2011}
   \maketitle
%

During the analysis of the HST/STIS observations of the NaI line in the atmosphere of the exoplanet HD189733b (Huitson et al., submitted), an error was detected in the code used for the HD209458b data analysis (Vidal-Madjar et al.~2011). In the conversion of the absorption depth into altitude, the altitudes were systematically overestimated.
Following the correction of this mistake, the atmospheric parameters were re-evaluated.
  

The change in the altitude evaluations do not significantly affect the temperature estimations because the spectral domains covered by each temperature step are relatively narrow. However, the physical sizes of the atmospheric layers are now smaller, and if we keep the same layers some are now smaller than the atmospheric scale heights.
Therefore, to properly correct the Table~1 of the Vidal-Madjar et~al.~(2011) paper, the number of considered layers has been decreased to four. The revised Table~1 and the resulting $T$-$P$ profile (Fig.~9 of the original paper) have been updated (hereafter Table~1 and Fig.~1).

  \begin{table*}[h]
\caption{\label{t7}Evaluated atmospheric parameters within adjusted atmospheric layers in order to have $H/\Delta z > 1$.}
\centering
\begin{tabular}{lcccccc}
\hline\hline
$Z_{\mathrm{min}}$ & $Z_{\mathrm{min}}$ & $T \pm \Delta T$ & $H$ & $H/\Delta z$ & $P_{\mathrm{max}}$ & $P_{\mathrm{min}}$ \\
(km) & (km) & (K) & (km) & & (bar) & (bar) \\ 
\hline
\vspace{1mm}
0 & 300 & $365 \pm 200$	& 133 & $2.2^{+2.8}_{-0.8}$ & $3.0 \times 10^{-3}$ & $3.2 \times 10^{-4}$ \\
\vspace{1mm}
300 & 1000 & $500 \pm 200$ & 183 & $3.8^{+2.6}_{-1.1}$ & $3.2 \times 10^{-4}$ & $6.9 \times 10^{-6}$ \\
\vspace{1mm}
1000 & 2350 & $1400 \pm 200$ & 511 & $2.7 \pm 0.4$ & $6.9 \times 10^{-6}$ & $4.8 \times 10^{-7}$ \\
\vspace{1mm}
2350	 & 4450 & $3600 \pm 1400$ & 1315 & $1.6^{+1.0}_{-0.5}$ & $4.8 \times 10^{-7}$ & $9.7 \times 10^{-8}$ \\
\hline
\end{tabular}
\end{table*}

The most important update concerns the pressure estimates. Because the overall atmospheric layer covered by the observations is smaller, in hydrostatic equilibrium the overall decrease in pressure with altitude is less.

The resulting new $T$-$P$ profile (Fig.~9 in Vidal-Madjar et al.~2011) is shown in Fig.~1. 

   \begin{figure*}
   \centering
  \includegraphics[width=14cm]{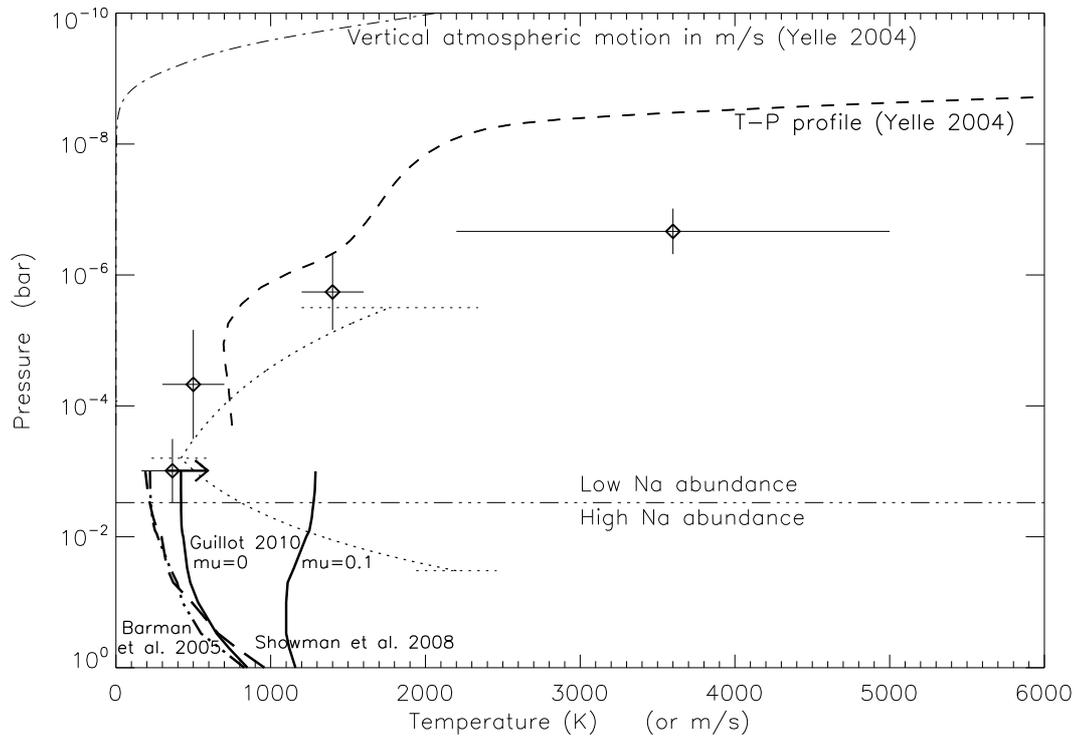}
   \caption{Plot of the new $T$-$P$ profile (diamonds). The symbols are the same as in Fig.~9 of Vidal-Madjar et al.\ (2011). The vertical error bars show the altitude regime over which we fit each temperature. The new $T$-$P$ profile remains consistent with the $T$-$P$ profile obtained by Sing et al.\ (2008) using a parametric fit of the same data set (dotted line). }
              \label{fig_tp}%
    \end{figure*}
    
The absorption profile as a function of bandwidth (Fig.~2 in Vidal-Madjar et al.\ 2011) shows a plateau interpreted in terms of a sodium abundance drop, likely because of ionization. The abundance drop of a factor of 4 remains unchanged, but the ionization pressure is now estimated to be around $7 \times 10^{-6}$\,bar rather than around $3 \times 10^{-5}$\,bar.

Most important, we confirm the detection of the thermosphere in the atmosphere of HD209458b which is not affected by the error in the original paper. We find that the thermobase corresponds to a rise in temperature at more than 2200\,K above the pressure level of $5 \times 10^{-7}$\,bar.
The new profile shows a more pronounced discrepancy with the prediction of Yelle (2004). The temperature rise in the thermosphere is now found at higher pressures than in the theoretical $T$-$P$ profile.

\section{Conclusion}

The results of our analysis favor the sodium condensation scenario to explain the deficiency of sodium observed in the line core and the possible presence of an Na ionization layer just below the base of the thermosphere.

Our temperature profile presents the following patterns:

\begin{itemize}

\item a rise from 365~K to 1400~K over nearly 1000~km (about 6 scale heights) in the pressure range of $3 \times 10^{-3}$ - $7 \times 10^{-6}$~bar with a relatively long isothermal layer at about 500~K over three scale heights over the $3 \times 10^{-4}$ - $7 \times 10^{-6}$~bar range;
\item a thick layer with a nearly constant temperature at about 1400~K. This layer has a thickness of about 1350~km corresponding to nearly three scale heights for a pressure range of $7 \times 10^{-6}$ - $5 \times 10^{-7}$~bar;
\item a further temperature rise is observed at higher altitudes, up to 3600~K at the highest altitudes, at nearly 4500~km above the reference level, corresponding to a pressure of about $10^{-7}$~bar.

\end{itemize}


\end{document}